\documentclass{llncs}

\usepackage{graphicx}
\usepackage[nolist]{acronym}
\usepackage{tikz}
\usepackage[hyphens]{url}
\usepackage{hyperref}
\usepackage[ colorinlistoftodos, textsize=tiny, ngerman ]{todonotes}

\begin{document}

\begin{acronym}
	\acro{CDN}{content distribution network}
	\acro{CMS}{content management system}
	\acro{DKIM}{DomainKey Identified Mail}
	\acro{DMARC}{Domain-based Message Authentication, Reporting and Conformance}
	\acro{DNS}{Domain Name System}
	\acro{DoS}{denial of service}
	\acro{DPA}{data protection authority}
	\acrodefplural{DPA}{data protection authorities}
	\acro{DPO}{data protection officer}
	\acro{GDPR}{General Data Protection Regulation}
	\acro{HSTS}{HTTP Strict Transport Security}
	\acro{NGO}{non-governmental organization}
	\acro{SPF}{Sender Policy Framework}
\end{acronym}

\hyphenation{Privacy-Score}

\title{PrivacyScore: Improving Privacy and Security via Crowd-Sourced Benchmarks of Websites\thanks{A German version of this paper with a more detailed discussion of the legal considerations is available at \cite{Maass2017RuT}. This is the authors' version. The final publication is available at link.springer.com via \href{http://dx.doi.org/10.1007/978-3-319-67280-9_10}{dx.doi.org/10.1007/978-3-319-67280-9\_10}}}

\author{Max Maass\inst{1}, Pascal Wichmann\inst{2}, Henning Prid\"ohl\inst{2}, Dominik Herrmann\inst{2}}

\institute{Technische Universit\"at Darmstadt, Secure Mobile Networking Lab, Germany \texttt{mmaass@seemoo.tu-darmstadt.de} \and Universit\"at Hamburg, Security in Distributed Systems Group, Germany \texttt{\{wichmann|pridoehl|herrmann\}@informatik.uni-hamburg.de}}
\maketitle

\begin{abstract}
Website owners make conscious and unconscious decisions that affect their users, potentially exposing them to privacy and security risks in the process.
In this paper we introduce PrivacyScore, an automated website scanning portal that allows anyone to benchmark security and privacy features of multiple websites. 
In contrast to existing projects, the checks implemented in PrivacyScore cover a wider range of potential privacy and security issues.
Furthermore, users can control the ranking and analysis methodology. Therefore, PrivacyScore can also be used by data protection authorities to perform regularly scheduled compliance checks. In the long term we hope that the transparency resulting from the published assessments creates an incentive for website owners to improve their sites. 
The public availability of a first version of PrivacyScore was announced at the ENISA Annual Privacy Forum in June 2017.

\end{abstract}


\section{Introduction}\label{sec:Introduction}
Setting up and running a website requires expert knowledge and a substantial amount of resources. Software systems have to be configured correctly, kept up-to-date and secured against attacks that are discovered during the lifetime of a site. %
The continuing flow of reports about security incidents on major websites indicates that many website operators are incapable of maintaining a sufficient level of security.

Vulnerabilities resulting from mistakes and negligence of website operators constitute a privacy risk for users. For instance, insecurely configured transport encryption may be broken by eavesdroppers, and sensitive data on web servers may be stolen by criminals. %

However, the privacy of users may also be under attack by the website operators themselves. Modern web design relies on third-party services: analytics services provide insights about visitors and ad networks generate revenue. 
Site owners commonly choose privacy-infringing third-party services, even though many typical requirements can be met with privacy-friendly alternatives, e.\,g., by running a local analytics tool such as Piwik \cite{piwik}.

The security and privacy risks resulting from (un)conscious decisions of site operators are complex and elusive. There is no straightforward way for end users to determine whether a site takes security seriously and whether it respects their privacy.

Existing website scanning services do not give a comprehensive impression of security and privacy features of a site.
First, these services are mostly geared towards skilled administrators to assist with self-assessment. 
Secondly, most scanners focus on a very specific area: the properties of the encrypted connections from the browser to the web server.
Thirdly, the interface of existing scanners makes it difficult to compare the results of different sites side-by-side.
As a result, site operators have little incentive to improve security and privacy on their website beyond the status quo.

Our project PrivacyScore aims to fill this gap. Building on existing work, and in cooperation with \acp{DPA} and data protection \acp{NGO}, we are currently designing and implementing the \textbf{PrivacyScore Benchmarking Portal}\footnote{Available online at \href{https://privacyscore.org/}{https://privacyscore.org/}.} to assess both security and privacy measures of websites. The public availability of a first version of PrivacyScore was announced at the ENISA Annual Privacy Forum in June 2017.

The target audience of PrivacyScore are \emph{end users} who want to know how a particular website ranks in its peer group, \emph{researchers} who want to perform studies about security features of websites, and \emph{\acp{NGO} and \acp{DPA}} that want to check the compliance of websites with data protection laws.

Our platform offers a scanning infrastructure and is built on the crowd-sourcing paradigm. 
It offers the following distinctive features:

\begin{itemize}
	\item Users can set up crowd-sourced lists of websites that belong to a certain category (e.\,g., all public schools in France). These site lists and their corresponding rankings are publicly visible and updated automatically on a regular basis to create an incentive for operators to improve their security and privacy compared to their peers or competitors. The results contain actionable advice for operators who want to improve their rating.
	\item Both list creators and ordinary users can adapt the rating and ranking of sites via customized ranking schemes.
	This allows list creators to create benchmarks that highlight particular checks, while giving users the opportunity to tailor the ranking to their personal privacy preferences.
	\item PrivacyScore is released under an open-source license (GPLv3 or later) and the data collected on the platform is made available to the public in human-readable and machine-readable form.
\end{itemize}

The rest of the paper is structured as follows: After having reviewed related work in Sect.~\ref{sec:relatedwork}, we present the main features of PrivacyScore in Sect.~\ref{sec:overview}. Section~\ref{sec:checks} provides details on our security and privacy checks, while Sect.~\ref{sec:implementation} outlines the implementation. Finally, we discuss legal and ethical considerations in Sect.~\ref{sec:considerations} before we conclude in Sect.~\ref{sec:conclusion}.

\section{Related Work}
\label{sec:relatedwork}
Most existing scanning services focus on \emph{security issues} which may allow malicious adversaries to compromise a web server or to eavesdrop on encrypted traffic. Prominent examples are the SSL scanners by Qualys \cite{qualys}, High-Tech Bridge \cite{htbridge}, and Mozilla \cite{mozilla}. Some of these services also check for the presence of relatively new HTTP headers \cite{securityheaders}, which protect against selected attacks.

On the other hand, there are only very few scanning services that focus on \emph{privacy issues}, i.\,e., design decisions allowing site owners or third parties to track users. A popular service is ``Webbkoll'' \cite{webbkoll}, which is offered by \emph{dataskydd.net}, a Swedish non-profit data protection organization.

Besides scanning services that allow users to scan arbitrary websites, there have been several efforts to scan pre-defined lists of websites and IP addresses. For instance, Helme regularly publishes a dataset containing an analysis of the HTTP security headers for the ``Alexa Top 1 Million'' list \cite{scotthelme}. Many more scans are available in the \emph{scans.io} repository.

Furthermore, there are numerous scientific studies that have analyzed security (e.\,g., \cite{HolzAMKW16,MayerZSH16}) and privacy aspects (e.\,g., \cite{Englehardt2016census}) of popular websites. These studies provide insights about the overall state of privacy and the adoption of security technologies on the web. However, as they typically focus on aggregate statistics, they do not create strong incentives for individual site owners to improve. Moreover, they rarely provide sector-specific insights (\cite{Englehardt2016census} is one of the few exceptions). And finally, as the published data is typically not updated after the publication, the results become outdated rather quickly. This is also true for the 2016 municipality survey \cite{kommuner} of \emph{dataskydd.net}, that inspired us to create PrivacyScore.

More distantly related to our work are browser add-ons such as Lightbeam \cite{lightbeam} and EFF's Privacy Badger \cite{badger}. These tools analyze visited sites on the fly and allow users to block dedicated tracking services. However, some add-ons have been shown to track their own users \cite{Starov2017,wot}.

With the existing projects and tools checking sites for legal compliance and comparing multiple sites are tedious processes. At the moment these tasks involve substantial amounts of manual work, because various results have to be obtained from multiple sources. With PrivacyScore we want to unify and simplify this process so that it becomes easily repeatable.

\section{PrivacyScore Overview}
\label{sec:overview}
\begin{figure}
\centering
	\includegraphics[width=1\columnwidth]{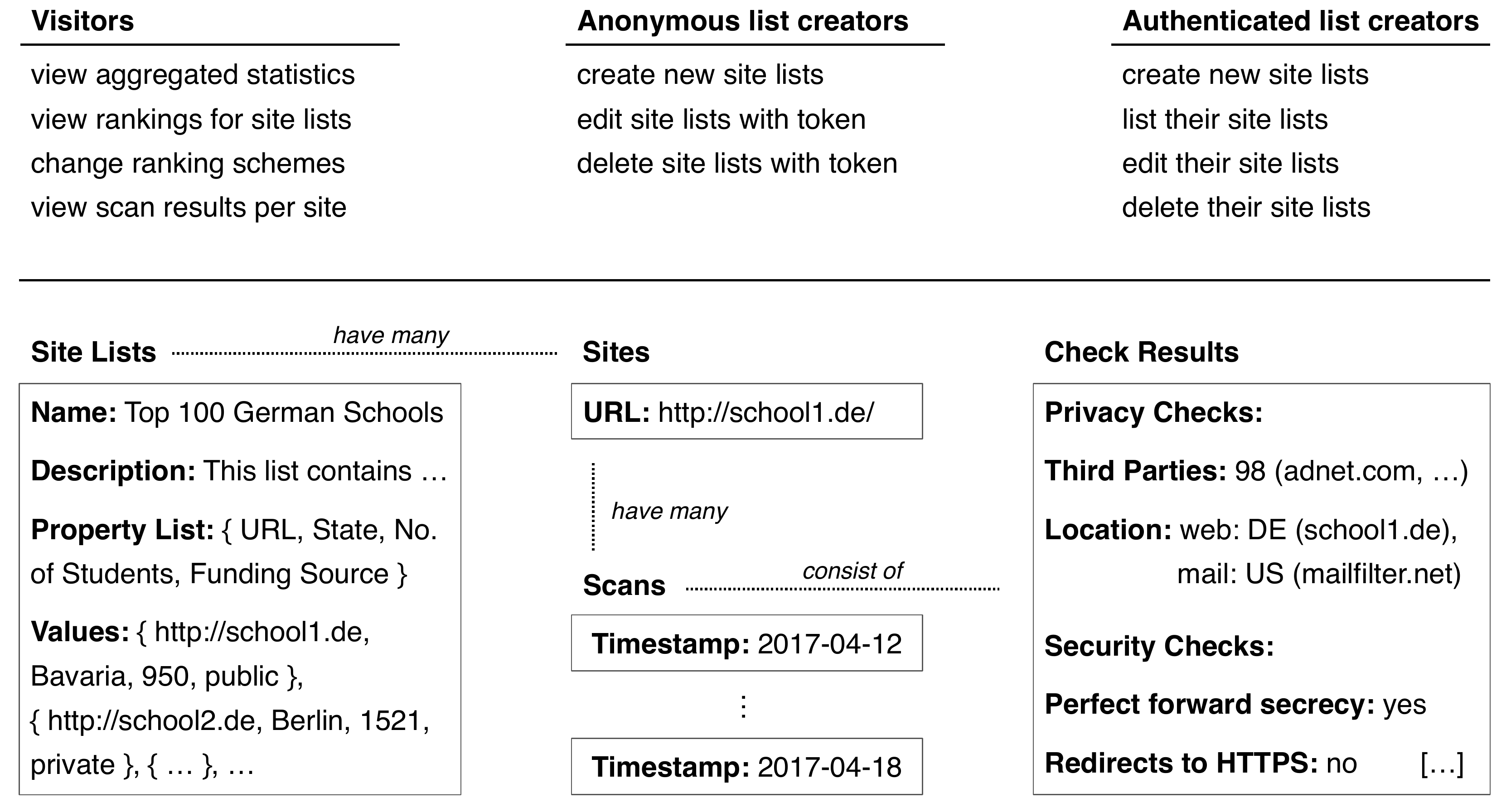} %
	\caption{\label{fig:usecases}Use cases and data structures}
\end{figure}

In this section we describe the main features of the PrivacyScore platform. Figure~\ref{fig:usecases} provides an overview of the use cases and data structures.

\subsection{Main Use Cases}
Like other website scanning services, PrivacyScore allows users to submit URLs of websites which are then analyzed in terms of security and privacy features. While one-off single-site scans are supported, the main purpose of PrivacyScore is the creation and publication of site lists that comprise multiple websites that share a common feature.
For instance, a list could consist of the websites of all schools of a country, of major newspapers, or popular health portals.

Users can browse the database of existing site lists (using tags and a full-text search) or create new site lists. New site lists can be created by anyone at any time. A site list is created by supplying a \emph{list of website URLs}. In addition, the creator of a site list can supply metadata such as a title for the list, a description of the methodology used to select the URLs, and a set of tags.
Once all relevant data has been entered (or uploaded using a CSV file), it is submitted to the scanning engine, which retrieves each site and gathers a number of facts using multiple scan modules. The gathered facts are evaluated by checks that assess specific security and privacy properties (cf. Sect.~\ref{sec:checks}).
The results can be displayed at varying levels of detail, from aggregate statistics over a tabular ranking of all sites to detailed results for each individual check and site.

Site lists and results are retained in the system. All lists are \emph{re-scanned on a regular basis} by default in order to document when site operators make changes to their sites (site list creators can disable automatic re-scans).

The system supports both incidental and professional use.
Occasional users can create site lists or scan individual websites without registration.
They receive a randomly generated \emph{access token} that allows them to change and delete their site list at a later time.
Professional users like \acp{DPA} can create a \emph{user account} to manage their site lists without having to keep track of their access tokens.

\subsection{User-centric Results}

PrivacyScore has two features that enable users to create meaningful assessments: \emph{user-defined properties} and \emph{user-defined ranking schemes}.

\paragraph{User-defined Properties}

A set of properties defined by the creator of a list can be stored for each site within a list. These properties can provide additional insights. The set of properties and their initial values are supplied by the creator of a site list, but they can also be refined and updated at a later time.
In the example of school websites, the following properties might be of interest: location (federal state), number of students, and whether it is publicly or privately funded. When users view the results of the assessment, they can use the properties to perform comparisons like ``is there a difference between public and private schools?''.

\paragraph{User-defined Ranking Schemes}

The ranking of the scanned websites is obtained by aggregating the results of the individual checks using a user-defined ranking scheme. Most existing scanning services use predefined weights to model the fact that some vulnerabilities are more critical than others.
In contrast to existing scanning services, where the scanning service imposes its ranking methodology on all scans, PrivacyScore gives its users more control. Our user-defined ranking schemes make the results useful for different audiences. For instance, \acp{DPA} may solely be interested in features indicating non-compliance with data protection law, while privacy activists may want a more strict rating for the purpose of ``naming and shaming''  websites with excessive data collection practices.

List creators can choose from a set of pre-defined ranking schemes. Moreover, users who view a site list can switch to a different ranking scheme or define their own on the fly. Our framework allows to calculate (potentially weighted) total scores or to define different types of ratings, for instance based on school grades (e.\,g., A\textsuperscript{+} to F) or by color-coding the result (e.\,g., red, yellow, green).

In order to make the results more easily accessible, checks are organized in \emph{check groups}. A ranking scheme defines how checks are organized into groups, how each possible outcome of a check is supposed to be rated, and which importance is associated to each check and group.

\begin{figure}[t]
\centering
	\includegraphics[width=1\columnwidth]{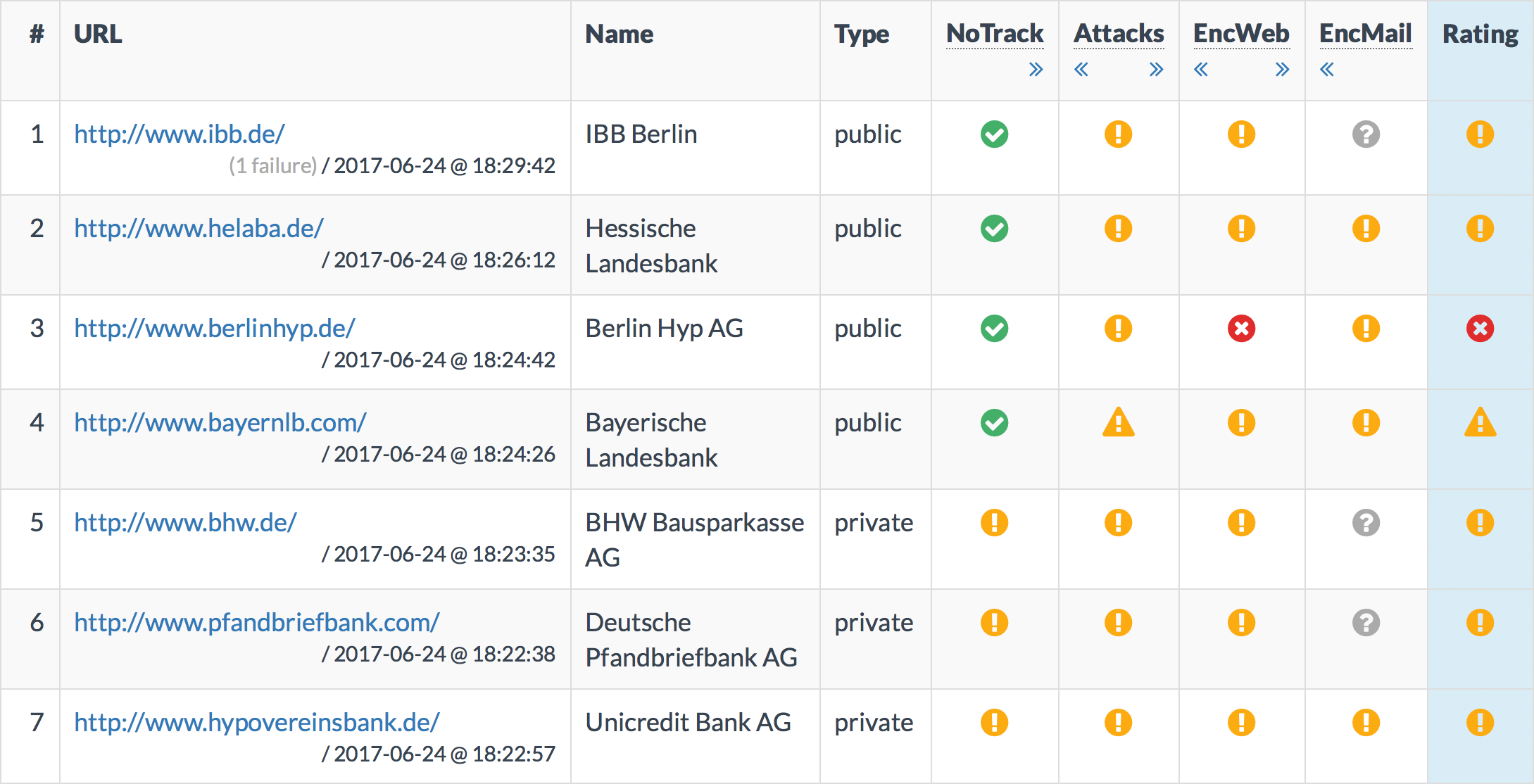} %
	\caption{\label{fig:ranking}Ranking for a site list that contains home pages of German banks}
\end{figure}

\paragraph{Rating and Ranking in the Beta}

In the following we explain the ranking scheme that is available at the start of the beta phase (cf. Fig.~\ref{fig:ranking}). This is subject to change and we will provide a more comprehensive description of the methodology at a later time.
For now the mapping of checks to check groups cannot be influenced by users and there are four check groups: \emph{NoTrack}, \emph{Attacks}, \emph{EncWeb}, and \emph{EncMail} (cf. Sect.~\ref{sec:checks}).

The currently implemented ranking scheme is based on the user-defined \emph{order of the check groups}. Users can manipulate the order according to their personal preferences.
Sites with a ``good'' (green) rating in the highest-priority check group (leftmost column) are pulled to the top of the table, followed by sites with a yellow and red rating in the first check group, respectively.
All sites with the same rating (color) in the first check group are further sorted according to their rating in the next check group -- and so on until all check groups have been considered.

\begin{figure}
\centering
	\includegraphics[width=1\columnwidth]{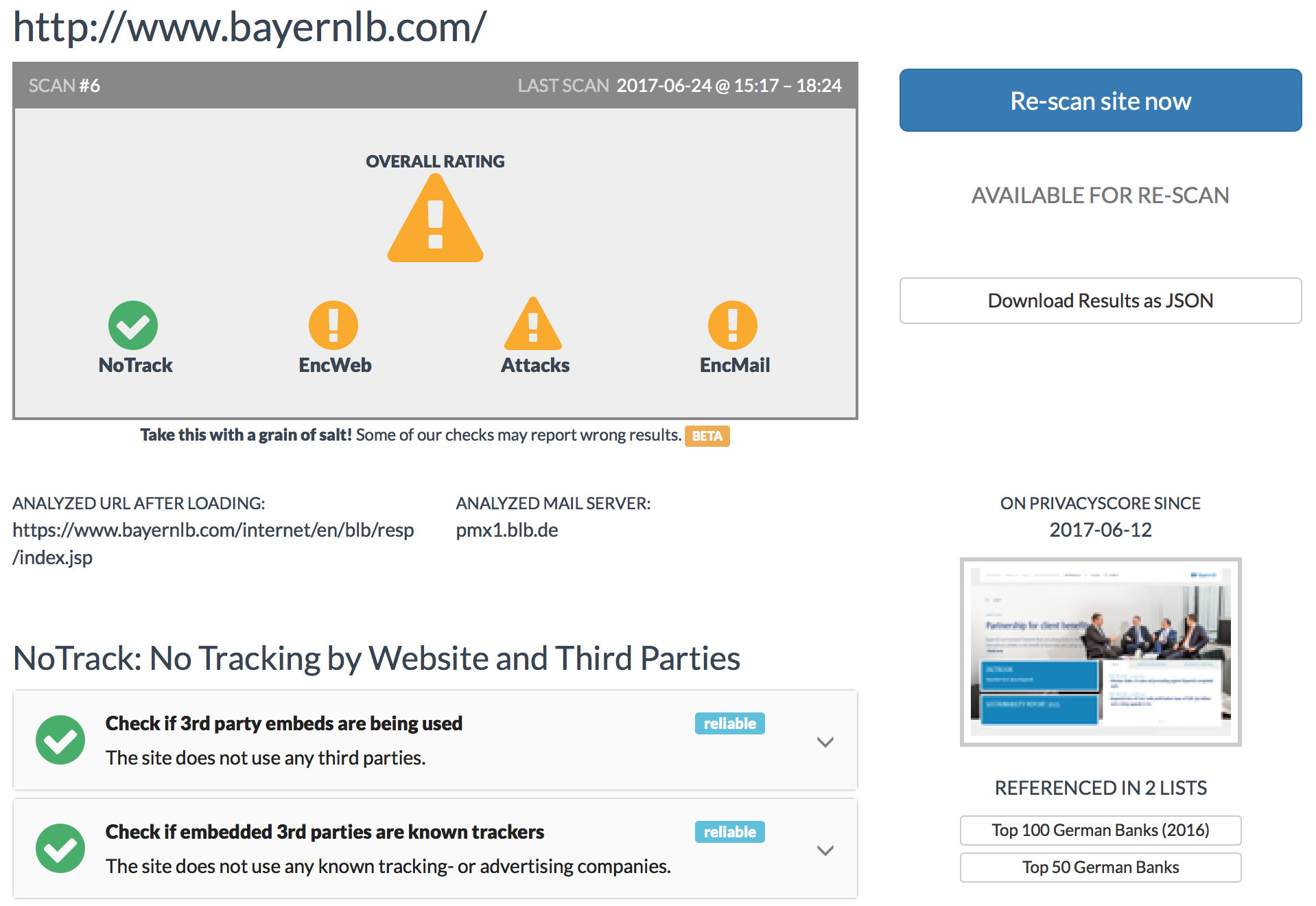} %
	\caption{\label{fig:details}Scan results for individual sites contain a list of all checks.}
\end{figure}

The rating of a site in a specific check group is determined by the checks belonging to that group. In the beta phase, a \emph{green} rating is obtained if all checks of a group succeed, while a \emph{red} rating is obtained if one of the critical checks within a group failed. Otherwise the site obtains a \emph{yellow} rating. The criteria for each check are documented on the PrivacyScore website (cf. Fig.~\ref{fig:details}).

Finally, the \emph{overall rating} of a site is given by the rating of the worst group. For instance, a site that has only green group ratings gets a green overall rating, while a site with at least one yellow or red group rating gets a yellow or red overall rating, respectively.

\subsection{Open Data versus Privacy}

We have designed PrivacyScore with the intention to improve transparency for end users by creating awareness for poor security and privacy practices of site operators. Therefore, all data that is generated on the platform is available publicly via an open RESTful API.

However, we recognize that some professional users have special privacy requirements. Therefore, all of the scan modules implemented on PrivacyScore run locally, i.\,e., the scanned URLs are not leaked to third parties. In addition, in order to support private investigations, site list can be marked as \emph{private}. This allows corporate users and \acp{DPA} to use PrivacyScore without disclosing this fact to the public. Furthermore, site lists can be deleted by the list creator or a system administrator upon request.

We also support users with even stricter privacy requirements, who can run their own (in-house) instance of PrivacyScore, as we release our source code under the GPLv3+ license.\footnote{See \href{https://github.com/privacyscore}{https://github.com/privacyscore}}

The open source nature also makes it easy to add additional scan modules, which are Python modules providing a specific interface. That way, scan modules can either be implemented directly in Python or just use a simple Python wrapper calling any external executable.

\section{Privacy and Security Checks}
\label{sec:checks}
PrivacyScore is designed to perform various checks on each website. In the following we describe our roadmap, i.\,e., the checks that we plan to support in PrivacyScore.
We distinguish two types of checks:
\emph{Security checks} analyze whether site operators follow best practices that protect against malicious attacks by outsiders. These checks are relevant because successful attacks may infringe the privacy of users. We also perform various \emph{privacy checks} that determine whether the owners of a site designed it in such a way that it infringes the privacy of users.

\subsection{Privacy Checks}
The privacy checks are reported in the \emph{NoTrack} check group in the beta phase (cf. Figs.~\ref{fig:ranking} and \ref{fig:details}).
The most prevalent privacy problem on modern websites is the plethora of tracking, analysis, and advertising services.
These services are usually embedded as JavaScript files that are retrieved from a web server run by the service provider (the so-called ``third party'').
The privacy and security implications of including third-party services in a website have been widely discussed \cite{Mayer12-thirdpartytracking}. Nevertheless, their use is ubiquitous on the modern web.

PrivacyScore enumerates the \emph{hostnames of third parties} that are included when a website is visited. The hostname of every third party is checked against a list of known advertisers and trackers (extracted from the EasyList \cite{easylist}) to determine which of them are trackers.

A common method for tracking users across multiple sites or over multiple visits to the same site are \emph{HTTP cookies}.
PrivacyScore measures how many cookies are being set, and how many of these are owned by third parties (which could use them for cross-page tracking of users).
A more modern technique for tracking and re-identifying users is \emph{browser fingerprinting}.
Here, characteristics of the browser (e.g., installed plugins and their version) and the device (e.g., available fonts and screen resolution) are being measured to compute a fingerprint.
Studies have shown that this fingerprint can often uniquely identify a browser \cite{Eckersley10-uniquebrowser,LaperdrixRB16}.
PrivacyScore will check whether the source code of a website contains known patterns of browser fingerprinting.

Furthermore, many websites rely on \acp{CDN} or load balancing services run by third parties, either for the site itself or in order to include software libraries such as jQuery. \acp{CDN} reduce page load times and improve scalability. However, they also pose a risk to the privacy and security of the users because the \ac{CDN} operators have full access to the unencrypted traffic. 
This was demonstrated by a recent vulnerability in the Cloudflare \ac{DoS} protection service, which exposed private information from thousands of websites \cite{Cloudbleed}.
Additionally, the \ac{CDN} companies themselves could track users on all websites that use their services.
Therefore, PrivacyScore checks whether the website itself or content from third parties is served from popular \acp{CDN}.

Finally, the \emph{geographic location of servers} may be of interest.
If the server is hosted under a different jurisdiction than the company itself, additional data protection rules may apply.
While the prevalence of \acp{CDN} with colocations in many countries makes determining the geographic location of servers and their associated jurisdictions complicated, a first approximation can be made using a GeoIP database.
PrivacyScore checks the location of the web-, mail-, and nameservers used by a website.

\subsection{Security Checks}

Insecure websites are more likely to suffer from breaches, unintentional data leaks, or data monitoring by rogue hotspots or ISPs.
PrivacyScore includes checks for security issues that may indicate privacy problems.

The most straightforward security feature a website can offer is to encrypt HTTP connections using TLS.
PrivacyScore checks if the website offers TLS, and if yes, whether unencrypted connection attempts are automatically forwarded to the HTTPS version of the site.
It also checks if the website follows established best practices for TLS deployment, if it is vulnerable to known attacks (such as Heartbleed and POODLE), and if the website contains unencrypted content on an encrypted page (\emph{mixed content}). These checks are part of the \emph{EncWeb} check group.
Similar checks are run for the primary \emph{mail server} that is listed in the MX domain record of the site, if one exists (\emph{EncMail} check group).

We also check whether a website sets \emph{HTTP security headers} like Content-Security-Policy and X-XSS-Protection that protect users from certain attacks. Together with the checks for unintended information leaks (described in the next paragraph) these checks make up the \emph{Attacks} check group.

The checks for unintended information leaks try to retrieve content from various well-known locations on a web server. Leaks may occur at the locations \emph{/server-status/} and \emph{/server-info/}, where the Apache server software publishes details about recently served requests (including source IP address) and the current load. We also check whether the operators have forgotten to remove frequently used test scripts (\emph{test.php} and \emph{phpinfo.php}) in the root directory of the server. These scripts may disclose the version and configuration of the server. We also check for the presence of \emph{/.git/} and \emph{/.svn/} directories. Operators that use version control systems to manage their sites should prevent unauthorized access to these locations. Otherwise, adversaries could retrieve the data stored in these locations to inspect the source code of server-side scripts, which may contain sensitive information such as database access credentials. Finally, we check for the presence of \emph{core dumps}, which are located at \emph{/core} and contain the memory of a process at the time it crashed. Core dumps can contain private information and should not be exposed publicly.

Another aspect that is often overlooked in practice is the security of the underlying \ac{DNS} records.
If the DNS entries can be compromised, sophisticated phishing attacks may be possible, regardless of all other security features of a website.
Accordingly, PrivacyScore checks if the DNS records of a site are protected with \emph{DNSSEC}.
PrivacyScore also checks if the mail server of a site uses state-of-the-art authentication techniques by consulting \ac{SPF} and \ac{DMARC} records in the \ac{DNS} \cite{RFC7208,RFC7489}.

The final building block in website security is keeping the software up to date. When PrivacyScore retrieves a website it searches for \emph{outdated software} by looking at version banners in headers sent by a server (``Server'' HTTP header, SMTP version banner string). It also tries to detect the version of the \ac{CMS} that is used to build the website (``generator'' attribute in HTML, and potentially file fingerprinting). Finally, PrivacyScore attempts to detect outdated client-side JavaScript libraries, which has been shown to be a surprisingly prevalent problem today \cite{Lauinger2017}.

\subsection{Incentives and Actionable Advice for Operators and Users}

PrivacyScore increases the visibility of security and privacy issues found on the scanned websites. However, we also want to increase the incentive for operators to improve their systems. Therefore, we plan to enrich the checks with explanations of the resulting security and privacy risks.
For instance, a missing or incorrectly set HSTS header means that malicious Wi-Fi access points can eavesdrop on traffic by performing an SSL stripping attack \cite{sslstrip}.

However, besides creating an incentive for site owners, our advice may also help attackers. 
For instance, we could create a tangible illustration of the risks resulting from outdated software by listing all relevant CVE entries \cite{cve} and reporting whether ready-to-run exploit modules for the Metasploit framework \cite{metasploit} are available. This example demonstrates that legal and ethical ramifications have to be considered to find an acceptable trade-off (cf. Sect.~\ref{sec:considerations}).

Furthermore, we want to support operators by offering actionable advice.
This includes suggesting alternative software (e.g., using a self-hosted analytics platform like Piwik \cite{piwik} instead of Google Analytics) and recommending specific configurations for popular operating systems, web servers, and \acp{CMS}.

Finally, we will support end users in protecting their privacy by suggesting protection strategies like installing browser add-ons such as PrivacyBadger \cite{badger} or uBlock Origin \cite{ublockorigin} that block advertisements and trackers.

\section{Implementation}
\label{sec:implementation}
\begin{figure}[t]
	\centering{
		\resizebox{0.95\textwidth}{!}{\input{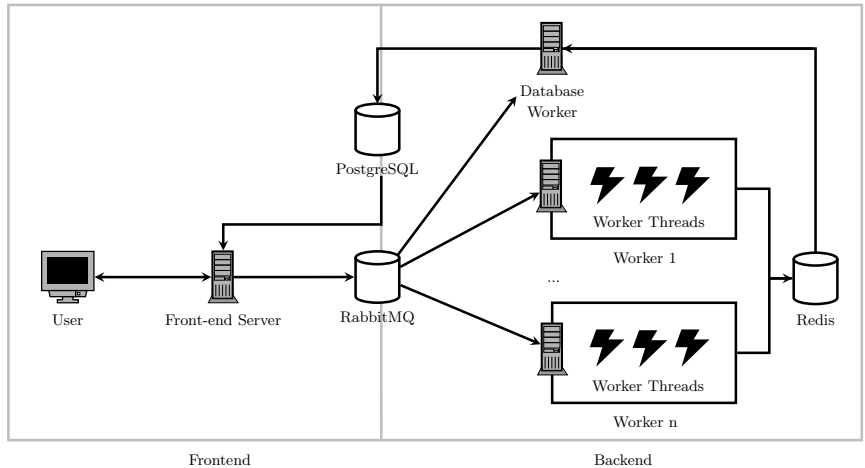}}}
	\caption{System Architecture of PrivacyScore}
	\label{fig:architecture}
\end{figure}

At the time of writing PrivacyScore is in a public beta phase and under active development (the benchmarking portal is available at \href{https://privacyscore.org/}{https://privacyscore.org/}).
We are implementing the system using a multi-tier software architecture (cf. Fig.~\ref{fig:architecture}).
Data is collected by a number of dedicated scanning machines (workers), each running a scanning system implemented in Python.
Scanning jobs are managed using the distributed task queue Celery \cite{Celery}.

This architecture allows us to scan multiple websites concurrently and to delegate individual scan modules to different machines, making the system horizontally scalable. New machines can be dynamically added to decrease the time that is needed to collect the data for a scan of a site.

New scanning tasks are queued by the front-end server (written in Python), while the results are aggregated and stored in a PostgreSQL database by a back-end server.
For security reasons, the database is not publicly accessible.  Instead, the back-end offers selective access to the data via a RESTful API.
The front-end views are implemented in Python using the Django web framework \cite{django}.

\paragraph{Implemented Checks}

At the time of writing, we have implemented four scan modules.
The first scan module uses the \emph{OpenWPM} \cite{Englehardt2016census} framework to collect information directly from a website by visiting it with a remote-controlled Firefox browser (we plan to support different browsers as well as browsers running on mobile operating systems in the future). OpenWPM allows us to observe requests to third-party hosts as well as the cookies they set. We also check whether the web server sent any HTTP security headers.

The second scan module analyzes the security of the encrypted connections to the web server and (if available) to the mail server of a website using the \emph{testssl.sh} script developed by Dirk Wetter \cite{testssl}.

The third scan module performs GeoIP lookups to determine the geographic location of the web and mail servers and the fourth scan module checks whether a website leaks information about its internal configuration.

The remaining checks mentioned in Sect.~\ref{sec:checks} are currently in development. Furthermore, only the supplied URLs (or the final URL, if the browser is redirected) are checked at the moment. In the future, we may instruct the browser to click on a few random internal links to get a more comprehensive picture of each site.

\section{Legal and Ethical Considerations}
\label{sec:considerations}

Automated scanning of websites on the internet poses a number of legal and ethical questions. In the following we will mention important issues, which we are currently discussing with legal academics and practitioners.

\subsection{Legal Considerations}
In some jurisdictions performing an automated security scan of a website without permission granted by its operators may be illegal or constitute a breach of its terms of service. 

This issue can be tackled in various ways. We are currently operating with a manual ``opt out'' policy. Site operators can contact us and declare that they do not want their sites to be scanned in the future. The respective URLs will be added to a blacklist. However, the last known scan results remain visible on the PrivacyScore website, annotated with a note that the operator has disapproved further scanning.

In principle, the scanner could also be programmed to refrain from scanning a website if it encounters a ``Disallow'' entry in the file \emph{robots.txt} on the web server, which is a commonly used method for site operators to indicate to search engine robots that some or all parts of their site should not be indexed.
However, to the best of our knowledge, none of the existing website scanners respect the rules contained in the \emph{robots.txt} file. Furthermore, the Internet Archive has recently announced to ignore \emph{robots.txt} files on purpose \cite{archiveorg}.
A viable compromise for PrivacyScore might consist in delegating the question whether to honor statements from \emph{robots.txt} to the list creator.

As executing certain scan modules may violate local laws, operators planning to run an instance of PrivacyScore are advised to evaluate applicable laws before deploying a scanner in their country.
A more detailed analysis of the legal issues surrounding PrivacyScore is available in a German version of this paper \cite{Maass2017RuT}.

\subsection{Ethical Considerations}

It is our intent to help site operators and users, but not adversaries.  However, some of our checks and advice provide information that may be useful for attackers, i.\,e., PrivacyScore is a \emph{dual-use tool}. Therefore, we are carefully designing what checks we implement and how we present the results to the user.

Scanning a website generates traffic (typically less than one megabyte) and temporarily increases its CPU load. Therefore, each scan incurs some cost for the owners of a website.
We believe this to be ethically acceptable, because the purpose of having a publicly available website is to have visitors that download it. Furthermore, we consider the public good of providing assessments to outweigh the typically negligible costs of each scan for the website operators.

However, we have to ensure that we do not induce a critical load on a website, which might result in denial of service. This is especially relevant for the SSL checks, which generate a large number of connections. To prevent too frequent scans we implement rate-limiting controls ensuring that the scanned sites cannot be overloaded maliciously.

\section{Conclusion}
\label{sec:conclusion}

Running a secure and privacy-respecting website has become a demanding task. On the one hand, website owners must constantly adapt their security measures to novel threats.
On the other hand, they have to make the right decisions during the design and operation of their site in order to safeguard the privacy of their users.
Today, many sites offer poor security and privacy, either because site owners are unwilling to cover the additional costs or because privacy measures are in conflict with their business model.

We believe that some site owners would make more privacy-conscious decisions if they had an incentive to do so. 
Our project PrivacyScore aims to create such incentives by generating transparency and publicity.
It allows anyone to set up a benchmark for a peer group of websites. Security and privacy features of the sites in a group are automatically analyzed, resulting in a public, regularly updated, and user-controllable ranking.

PrivacyScore is open source software and we plan to release all collected datasets for research purposes. Besides running it as a public service, PrivacyScore can also be deployed in-house. 
We hope that it will become a useful tool for \acp{DPA} that are faced with the task of enforcing a large number of regulatory requirements specified in the \ac{GDPR}.

\paragraph{Acknowledgments}
\label{sec:Acknowledgments}

This work has been co-funded by the DFG as part of project C.1 within the RTG 2050 ``Privacy and Trust for Mobile Users''. The authors are grateful to Marvin Hebisch and Nico Vitt, who implemented a prototype, the attendants of the PET-CON 2017.1 workshop, and members of Digitalcourage~e.\,V. for their valuable suggestions.

\bibliography{bibliography}{}
\bibliographystyle{splncs03}

\end{document}